\documentclass{elsart} 
\usepackage{graphicx}
\usepackage{times}
\begin{document}
\begin{frontmatter}
\title{Structure determination of the 
$(3\sqrt{3}\times 3\sqrt{3})$ reconstructed 
$\alpha-$Al$_2$O$_3(0001)$}

\author[Grenoble,Ljubljana]{Igor Vilfan},
\author[Grenoble]{Thierry Deutsch}, 
\author[Grenoble]{Fr\'{e}d\'{e}ric Lan\c{c}on},
\author[Grenoble]{Gilles Renaud}        
\address[Grenoble]{D\'{e}partement de Recherche Fondamentale sur
      la Mati\`{e}re Condens\'{e}e, CEA-Grenoble, F-38054 Grenoble cedex 9,
      France}
\address[Ljubljana]{J. Stefan Institute, P.O.  Box 3000, SI-1001 Ljubljana,
      Slovenia\thanksref{per_add}}
\thanks[per_add]{Permanent address of the corresponding author.} 
\begin{abstract}
Grazing-incidence X-ray diffraction data are combined with 
energy-minimization calculations to analyse the atomic structure of 
the Al-rich $(3\sqrt{3}\times 3\sqrt{3})$R$30^\circ$
reconstructed surface of sapphire $\alpha-$Al$_2$O$_3(0001)$.
The experiments on the BM32 beamline of the ESRF provide the
non-integer-order diffraction intensities and, after Fourier transform,
an incomplete Patterson map. 
The computer simulations are implemented to obtain structural 
information from this map.
In the simulations, the interactions between the Al overlayer atoms 
were described with the Sutton-Chen potential and the interactions between
the overlayer and the sapphire substrate with a laterally modulated 
Lennard-Jones potential.
We have shown that the hexagonal reconstructed unit cell is composed of 
triangles where the two layers of Al adatoms are FCC(111) ordered
whereas between the triangles the stacking is FCC(001).
\end{abstract}
\end{frontmatter}

\section{Introduction}
The $(0001)$ surface of sapphire $\alpha-$Al$_2$O$_3$ is, 
because of its insulating character, an important 
substrate for very high frequency microelectronic devices.
Therefore it is of crucial importance to understand the stability
and structure of this surface at high temperatures.
In equilibrium at room temperature, a clean $(0001)$ surface is 
unreconstructed and is terminated with a single 
Al layer on top of the last oxygen layer \cite{GRBG97,DFN99,WSH00}.
Upon heating up to $\sim$1500 C in ultra-high vacuum (UHV),
oxygen evaporates from the crystal, leaving an Al rich and
reconstructed surface behind it \cite{FS70}.
Al atoms that are left on the sapphire surface are responsible 
for a sequence of surface reconstructions:
$(\sqrt{3}\times \sqrt{3})$R30$^\circ$ when sapphire is heated to 
about 1100 C, 
$(2\sqrt{3}\times 2\sqrt{3})$R30$^\circ$ at $~$1150 C,  
$(3\sqrt{3}\times 3\sqrt{3})$R30$^\circ$ at $~$1250 C and  finally  
$(\sqrt{31}\times \sqrt{31})$R$\pm9^\circ$ structure at $~$1350 C
\cite{R98}.
So far only the structure of the most stable, i.e., of the 
$(\sqrt{31}\times\sqrt{31})$R$\pm 9^\circ$ reconstruction has been
satisfactorily determined \cite{RVVB94,VLV97}.
Very recently, this structure has also been investigated by
Barth and Reichling with the
atomic-scale-resolution scanning force microscope \cite{BR01}.
No information on the real-space structures of other reconstructions  
is available so far. 

In this letter we concentrate on the  
$(3\sqrt{3}\times3\sqrt{3})$R$30^\circ$ reconstruction of the 
sapphire (0001) surface.  
Although this is a transient reconstruction, it was prepared in a 
very well defined state with large domain sizes \cite{R98,RBG}.

Our analysis is based on grazing-incidence X-ray diffraction (GIXD)
data where some surface diffraction intensities are hidden in the bulk 
Bragg peaks and the corresponding Patterson map is incomplete
\cite{YZ86}.
To obtain the real-space structure, therefore, we combined the
Patterson map with computer simulations using semi-empirical
potentials. 

\section{Experiment}
\begin{figure}
   \begin{center}
   \includegraphics{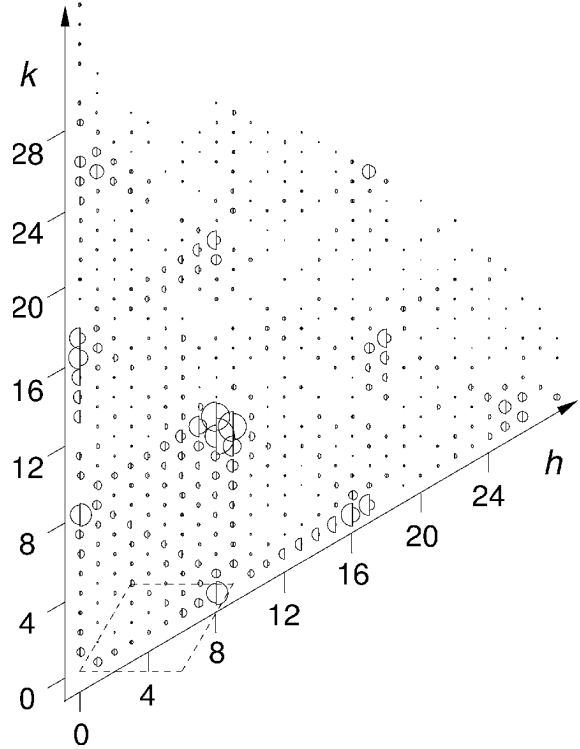}
   \end{center}
   \caption{Diffraction pattern of the 
   $(3\sqrt{3}\times3\sqrt{3})$R$30^\circ$ reconstructed surface
   of sapphire (0001). The areas of semi-circles are proportional
   to the intensity of diffraction spots. The right-hand 
   semi-circles are experimental results and the left-hand ones
   correspond to the structure proposed in this paper. 
   The in-plane bulk reflections are omitted from the graph.
   The dashed lines show the unreconstructed reciprocal cell.}
   \label{diff_p}
\end{figure}

The structure was investigated experimentally by grazing-incidence 
X-ray diffraction using the SUV set-up \cite{BRS99} 
of the BM32 beamline at the ESRF. 
This is equipped with a large UHV chamber 
($3\times 10^{-11}$ mbar base pressure) 
with RHEED, AES, mass analyser, a high temperature (up
to 1500 C) furnace and several deposition sources. 
The sample was prepared
by heating at ~1250 C for 20 min, which yielded a 
$(3\sqrt{3}\times3\sqrt{3})$
reconstruction of very high quality, as checked by \textit{in situ} RHEED, 
and negligible segregation of contaminants, as checked by AES. 
A 18 keV doubly focused monochromatic X-ray beam was used, 
with a $0.2(\rm{H}) \times 0.2(\rm{V})$ mm$^2$ size
and $0.5(\rm{H}) \times 0.1(\rm{V})$ mrad$^2$ convergence at the vertical 
sample location. 
The angular acceptance of the NaI detector was set to 
$20(\rm{H}) \times 4 (\rm{V})$ mrad$^2$. 
The incident angle was fixed at the critical angle for total external
reflection during the whole measurement. 
A total of 1790 in-plane reflections
were measured, showing $p6mm$ symmetry. 
These were integrated over azimuthal rotation and corrected for 
background, effective area, polarisation and Lorentz factors 
\cite{RGAV00} to extract the in-plane structure-factor amplitudes
$F_{hk0}$.  This reduced to $240$ inequivalent peaks. 
The experimental diffraction
pattern in $1/6$ of the reciprocal space is shown as right-hand
semi-circles in Fig. \ref{diff_p}, with their radii proportional 
to $F_{hk0}$. 
The $h$ and $k$ indices are in reciprocal lattice units of 
the $(3\sqrt{3}\times3\sqrt{3})$R$30^\circ$ surface reconstruction.
Since the phases of $F_{hk0}$ are unknown, 
we cannot directly determine the real-space density distribution.
In addition, the amplitudes at the bulk Bragg reflections are unknown
since the contribution of the surface reconstruction is hidden in the
bulk contribution. 
Therefore the Fourier transform of the experimental
$F_{hk0}^2$ gives an incomplete Patterson map,
i.e., an incomplete electron (atom) density-density correlation function. 
The incomplete Patterson map of the $(3\sqrt{3}\times3\sqrt{3})$R$30^\circ$ 
reconstructed sapphire (0001) is shown in Fig. \ref{patt}(a).

\section{Simulations} 
\begin{figure}
   \begin{center}
   \includegraphics{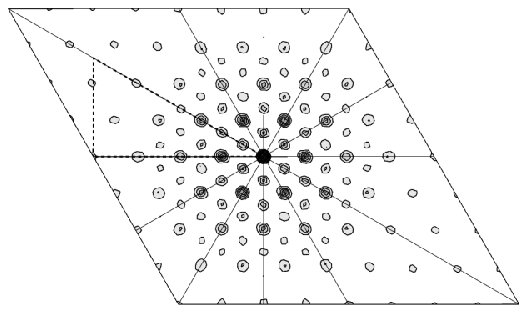}\hskip1cm
   \includegraphics{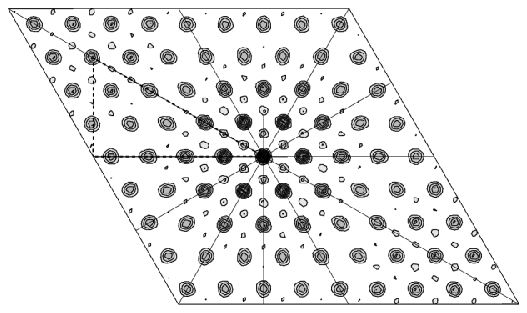}\\
  a\hskip6cm b\\
   \end{center}
   \caption{The Patterson map of one reconstructed 
    unit cell on sapphire.  (a) Incomplete Patterson map based
    on the experimental diffraction data only, 
    (b) Patterson map when the intensities calculated from the model 
    at the in-plane bulk reflections are added to the experimental 
    diffraction data.} 
   \label{patt}
\end{figure}

Although the Patterson map does not give the real-space 
structure, it is the starting point to build the real-space models.
It tells us that the surface atoms that contribute to the more 
compact regions of the Patterson map are ordered on a hexagonal lattice,
very similar to the ``ordered domains'' seen also in the 
$(\sqrt{31}\times \sqrt{31})$R$\pm9^\circ$ reconstruction of sapphire
\cite{RVVB94,VLV97}. 
Therefore we constructed the overlayer assuming that sapphire is
covered by two FCC(111) layers of Al with the nearest-neighbour
separation $\sim 4$ \% larger than the corresponding sapphire spacing.
The lattice misfit inevitably leads to perturbations in the
close-packed FCC(111) structure, creating a commensurate structure
composed of ``domains'', where the overlayer atoms have perfect
FCC(111) or hexagonal close-packed ordering, separated by a kind of 
``domain walls''. 
Whereas the atomic structure in the domain walls is disordered
in the $(\sqrt{31}\times \sqrt{31})$ reconstruction 
\cite{RVVB94,VLV97,BR01}, the Patterson map of the 
$(3\sqrt{3}\times3\sqrt{3})$ reconstruction in Fig. \ref{patt} 
suggests a much more ordered structure of domain walls.
We generated several initial structures with hexagonal close-packed
atom ordering in the domains with domain walls 
between them. 

Such reconstructions are too large to be investigated with 
\textit{ab initio} methods. 
Therefore we described the interactions between the Al overlayer atoms 
with the semi--empirical Sutton--Chen potential \cite{SC90,FS84} 
and the effect of the substrate on the overlayer with 
a phenomenological potential field with hexagonal symmetry. 
The energy minimization was described in detail in \cite{VLV97}.

The semi-empirical many-body Sutton-Chen potential 
\begin{equation}
   U_{\rm{SC}}  = {1\over 2}\sum_{i\ne j} \epsilon 
   \left( {a\over r_{ij}}\right)^n  - 
   \epsilon C \sum_{i} \sqrt{\rho_i}
   \label{Alpot}
\end{equation}
includes the core repulsion potential (first term) and  
the bonding energy mediated by the electrons (second term).
$r_{ij}$ is the separation between the atoms $i$ and $j$,    
$\rho_i$ is an effective local electron density at the site $i$:
\begin{equation}
    \rho_i = \sum_{j\ne i}\left( {a\over r_{ij}}\right)^m
\end{equation}
and $a$ is the lattice constant of an Al FCC crystal.
$\epsilon$ and $C$ are parameters of the model which, together 
with the exponents $n$ and $m$, determine the repulsive and cohesive
energies, respectively.  
We used the following values for the 
potential parameters of Al \cite{SC90}:
\begin{eqnarray*}
   m = 6, \quad n = 7, \quad \epsilon = 33.147\, \mbox{ meV},\quad
   C = 16.399.
\end{eqnarray*}
The potential was truncated continuously (with a fifth order 
polynomial) between $r/r_0 = 3.17$ and 
$3.32$
($r_0$ is the nearest-neighbour distance). 
In this way the interaction with 68  neighbours would be included 
if Al were perfectly ordered in two FCC(111) planes.

In general, the overlayers or adatoms are weakly bound to the 
substrate and well separated from it \cite{VJSS99}. 
In addition, the substrate is much stiffer than the Al overlayer, 
therefore the relaxation of the substrate caused by the overlayer 
was neglected in the simulations.
The substrate potential was expanded in a power series and 
only the six lowest-order terms were retained \cite{VLV97}, 
\begin{eqnarray}
   U_S =  U_{LJ}(z) 
   {U_{L} - \cos (\vec{k}_1 \cdot \vec{r}) \cos(\vec{k}_2 \cdot \vec{r})
     \cos(\vec{k}_3 \cdot \vec{r}) \over U_{L} - 1}\cr
\label{Usub}
\end{eqnarray} 
where            
\begin{eqnarray}
        \vec{k}_1 = {2\pi \over a_s} (0,1),  \quad
        \vec{k}_2 =  {\pi \over a_s} ( \sqrt{3}, -1 ), \quad
        \vec{k}_3 =  {\pi \over a_s} (-\sqrt{3}, -1 ) 
\end{eqnarray}
are the unit vectors in the plane of the surface and $a_s$ the substrate 
lattice constant.
The parameter $U_{L}$ controls the strength of the lateral modulation of 
the potential, and $U_{LJ}$,
\begin{equation}
    U_{LJ}(z) = U_0 \left[ \left({z_0\over z}\right)^{9} - 
                         2 \left({z_0\over z}\right)^{3}\right],
\label{U_LJ}
\end{equation}
its $z-$dependence, in the direction perpendicular to the surface.
Expression (\ref{U_LJ}) is the Lennard-Jones potential 
integrated over the semi-infinite substrate where
$U_0$ is the depth of the substrate potential and $z_0$ determines 
the position of the minimum and the width of the potential in the 
vertical direction. 
The potential has six equally deep minima in the surface unit cell,
to accommodate three overlayer atoms in each monolayer.
Thus, the overlayer--substrate potential is described by three 
\textit{variational parameters}: $U_L$, $U_0$ and $z_0$. 

The parameters $U_L$, $U_0$ and $z_0$ were varied to minimize
the difference between the experimental and calculated structure
factors \cite{R98}:
\begin{equation}
   \chi^2 = \frac{1}{N} \sum_{hk} \frac{\left(|F_{hk0}^{\rm{exp}}| -
   |F_{hk0}^{\rm{calc}}| \right)^2}{(\sigma_{hk0}^{\rm{exp}})^2}
\end{equation}
where the summation is over $N$ measured diffraction peaks, 
$\sigma_{hk0}^{\rm{exp}}$ are the experimental errors in 
$F_{hk0}^{\rm{exp}}$ whereas $F_{hk0}^{\rm{calc}}$ are the structure
factors of the trial structures.

\begin{figure}[h]
   \begin{center}
   \includegraphics{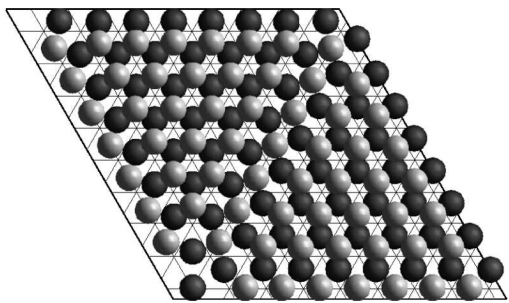}
   \end{center}
   \caption{The $(3\sqrt{3}\times3\sqrt{3})$R$30^\circ$ 
   reconstructed unit cell
   with 136 atoms (73 in the lower and 63 in the upper plane)
   in the calculated model.
   The darker spheres represent the atoms in the lower plane.
   The underlying triangular net represents the periodicity
   of the simulated substrate potential which has minima in
   the centres of the triangles.  }
   \label{str_ini}
\end{figure}

The initial structures with the number of atoms in reconstructed
unit cell varying between 134 and 140 were first tested for their 
stability in a reasonable range of substrate potentials
and then the parameters $U_L$, $U_0$ and $z_0$ were varied until
the minimal $\chi^2$ was reached.   
In this way we narrowed the possible initial structures
to one, having 136 surface 
atoms in the reconstructed unit cell, shown in Fig. \ref{str_ini}. 
This structure is very stable, it does not vary  after 
random atomic displacements up to 0.286 {\AA} and 
subsequent energy minimization.
The variational parameters in the minimum of $\chi^2$ 
($\chi^2 = 2.54$) 
are:  $U_0 = 0.28$ eV, $U_L = 23$  and $z_0 = 5.10$ \AA. 
The average Sutton-Chen potential energy per adatom is 
$E_{\rm{SC}} = - 3.045$ eV and the average adatom--substrate energy 
$E_S = -0.230$ eV, giving $E =  -3.275$ eV for the average total 
potential energy per adatom. 
$\chi^2$ varies by less than 10\% for $U_0$  varying 
between 0.16 and 0.70 eV. 

\section{Results and Discussion}
We would like to emphasize the strength of the 
present combined experimental-theoretical approach compared to
the standard surface X-ray diffraction analysis based only on
the Patterson-map analysis and $\chi^2$ minimization 
\cite{V00}. 

The first advantage is related to the weakness of the
standard method where the surface structure factors corresponding
to the locations of the bulk contributions are not included in
the Patterson analysis, thus leading to an incomplete
Patterson map \cite{YZ86}. 
Omission of some peaks from the diffraction data can introduce errors 
in the final atomic positions. 
In Fig.~\ref{patt} it is demonstrated that the peaks 
corresponding to $(h,k) \equiv (0,0)\pmod{9}$ have
a dramatic effect: when they are omitted, the incomplete 
Patterson function shows 8 atoms along the reconstruction primitive
vectors [Fig.~\ref{patt}(a)] instead of 9 that are 
retrieved if we complete the diffraction data with the model
values [Fig.~\ref{patt}(b)].
In case of  free FCC(111) stacked Al the lattice parameter 
is $\sim 4$~\% larger than the corresponding sapphire distance. 
In the present case, the misfit is $\sim 2 - 3$~\%, showing that 
the Al layer is strained by the underlying sapphire lattice.
Such overlayer yields significant intensities at the bulk peaks 
locations (corresponding to no expansion) as well as in the 
neighbouring peaks (12.5 \% expansion). 
A standard analysis would thus favour the second components in the
Al atom positions, hence probably over-estimating the expansion of
the Al overlayer. The present approach avoids this effect because
the bulk component is re-introduced via the interaction with the 
substrate, and thus the final structure would be closer to the
real one than with the standard analysis.

The second advantage of the present method is that it yields a
high-symmetry structure with negligible disorder which, we believe,
is close to the ``ideal'' structure of the $(3\sqrt{3}\times
3\sqrt{3})$ reconstruction while the experimental data 
correspond to a structure which may be substantially disordered. 

\begin{figure}
   \begin{center}
   \includegraphics[scale=1.5]{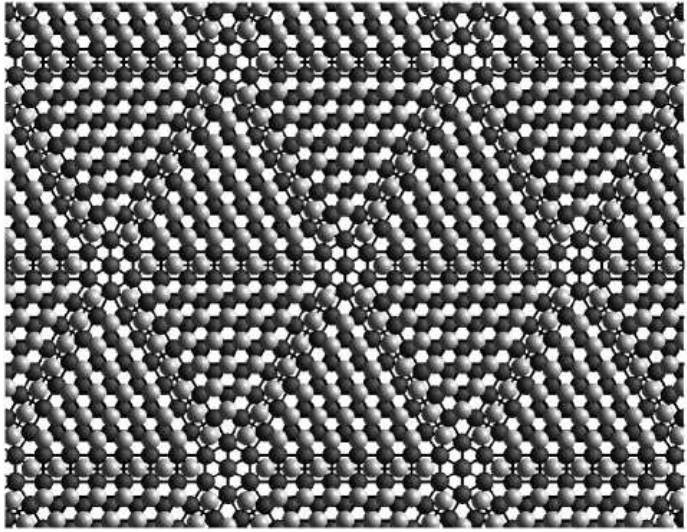}
   \end{center}
   \caption{The relaxed atomic configurations of the 
   $(3\sqrt{3}\times3\sqrt{3})$R$30^\circ$ 
   reconstruction. Clearly seen are the triangular domains with
   hexagonal [FCC(111)] structure separated by ``walls'' where the 
   structure is FCC(001). At the corners of the triangles the
   density is lower. 
   }   
   \label{str_final}
\end{figure}

The final, relaxed structure is shown in Fig. \ref{str_final} and 
in Fig. \ref{diff_p} we compare the calculated (left-hand semi-circles) 
diffraction pattern with the experiment (right-hand semi-circles).
Specific for this structure is a triangular pattern with the period
$3 \sqrt{3} a_s$. 
Inside these large triangles, the two Al planes are ordered hexagonally 
in FCC(111) planes. 
In the ``down''  triangle (upper left-hand side of Fig. \ref{str_ini})
the lower-layer Al atoms are located in the centres of the 
underlying small up-triangles, representing the substrate potential, 
and the upper-layer atoms in the centres of the small down-triangles. 
In the ``up'' triangle (lower right-hand side of Fig. \ref{str_ini}) 
the stacking is reversed. 
The two large triangles together, thus, form the rhombic reconstructed 
unit cell. 
Between the triangles, the lower-layer Al atoms are ordered in
squares; together with the atoms in the upper layer they form 
semi-octahedra with an  FCC(001) orientation. 
The semi-octahedral ordering is needed to compensate for the $\sim 4$\%
lattice misfit between the sapphire substrate and the Al overlayer.
In the corners of the large triangles we find small hexagonal vacancy
islands where several Al atoms are missing in the upper layer.
The average density of the overlayer atoms is $136/27 \approx 5$ 
Al atoms per unreconstructed surface unit cell and is surprisingly 
close to the density proposed for the $(\sqrt{31}\times \sqrt{31})$ 
reconstruction \cite{VLV97}. 
In the large triangles, the nearest-neighbour Al-Al spacing 
of the lower layer is between 2.82 and 2.84 \AA, giving a 
contraction of up to 1.5 \% with respect to bulk Al. The
spacing in the FCC(001) octahedra is similar, with a
contraction of up to 2 \%. 
In principle, our ground-state simulations predict a chiral structure
with $p6$ symmetry. 
The $p6mm$ symmetry is restored either by superimposing both chiral
orientations or by room-temperature thermal fluctuations since
the deviations from the mirror symmetry are small, of the order 
0.03 {\AA}. 
It is interesting to note that the structure of the 
$(3\sqrt{3}\times3\sqrt{3})$ reconstruction, as proposed in this paper, 
is much less disordered than the final, 
$(\sqrt{31}\times \sqrt{31})$ reconstruction \cite{VLV97,BR01}.

The difference between the experimental and simulated diffraction
patterns in Fig. \ref{diff_p} might have two origins.
First, the adatom-adatom and the adatom-substrate potentials
were treated in an approximate way, and second, there can be some 
disorder in the surfaces.
Indeed, experimentally, the reconstruction with $3\sqrt{3}$
symmetry can be prepared in very different states of order
(as checked by electron or X-ray diffraction), depending on the
exact annealing temperature and duration. 
A slight variation from the
best conditions leads to a high level of background scattering as
well as to different relative ratios of the structure factors.
A minimization of $\chi^2$ with respect to the Debye-Waller
factor using the ROD program \cite{V00}, for instance,  and keeping 
all the atomic coordinates fixed improves the $\chi^2$ value to 
$\chi^2 = 2.05$. 
The corresponding Debye-Waller factor increases 
to $\sigma = 0.25$~{\AA}, indicating that some kind of disorder (i.e., 
variation of the atomic structure between different reconstruction 
cells) is present on the prepared surface. 
In contrast, a standard minimization of $\chi^2$ with the ROD 
program varying the atomic positions 
yields a much better agreement  ($\chi^2$ as low as 0.7!) 
but for very disordered and unphysical structures.

To conclude, on the basis of our GIXD experiments combined with the  
energy minimization using effective potentials, we propose  
a real-space structure of the 
$(3\sqrt{3}\times3\sqrt{3})$ reconstruction of Al$_2$O$_3(0001)$.
We have shown that the combination of GIXD experiments with the
simulations based on energy-minimization is a powerful tool
to predict and understand the large unit cell reconstructions,
in particular on oxides or insulators, where atomic-scale 
information from scanning force microscopy is still extremely 
scarce.

\section*{Acknowledgement}
I.V. would like to acknowledge the support of the UFR de Physique, 
Universit{\'{e}} Joseph Fourier in Grenoble and the hospitality of 
the D{\'{e}}partement de Recherche Fondamentale sur la Mati\`{e}re 
Condens{\'{e}}e.
The comments by Clemens Barth are gratefully acknowledged.

\end{document}